\documentclass[sigconf,screen]{acmart}
\acmConference[ESEC/FSE 2023]{The 31st ACM Joint European Software Engineering Conference and Symposium on the Foundations of Software Engineering}{11 - 17 November, 2023}{San Francisco, USA}
\setcopyright{rightsretained}

\usepackage{code}
\usepackage{graphicx}
\usepackage{balance}
\usepackage{multirow}
\usepackage{multicol}
\usepackage{listings}
\usepackage{booktabs}
\usepackage{subcaption}
\usepackage[normalem]{ulem}
\usepackage{cleveref}
\usepackage{xspace}

\begin{document}
\acmPrice{15.00}
\acmDOI{10.1145/3611643.3616289}
\acmYear{2023}
\copyrightyear{2023}
\acmSubmissionID{fse23main-p415-p}
\acmISBN{979-8-4007-0327-0/23/12}
\acmConference[ESEC/FSE '23]{Proceedings of the 31st ACM Joint European Software Engineering Conference and Symposium on the Foundations of Software Engineering}{December 3--9, 2023}{San Francisco, CA, USA}
\acmBooktitle{Proceedings of the 31st ACM Joint European Software Engineering Conference and Symposium on the Foundations of Software Engineering (ESEC/FSE '23), December 3--9, 2023, San Francisco, CA, USA}
\received{2023-02-02}
\received[accepted]{2023-07-27}

\title{Contextual Predictive Mutation Testing}

\author{Kush Jain}
\affiliation{\institution{Carnegie Mellon University}
\country{United States}}

\author{Uri Alon}
\affiliation{\institution{Carnegie Mellon University}
\country{United States}}

\author{Alex Groce}
\affiliation{\institution{Northern Arizona University}
\country{United States}}

\author{Claire Le Goues}
\affiliation{\institution{Carnegie Mellon University}
\country{United States}}


\newcommand{\mr}[2]{\multirow{#1}{*}{#2}}
\newcommand{\mc}[3]{\multicolumn{#1}{#2}{#3}}

\newcommand{\clg}[1]{\textcolor{blue}{#1}}
\newcommand{\adg}[1]{\textcolor{purple}{#1}}
\newcommand{\kj}[1]{\textcolor{olive}{\textbf{\small[Kush: #1]}}}
\newcommand{\todo}[1]{\textcolor{red}{todo: #1}}
\newcommand{\uri}[1]{\textcolor{magenta}{\textbf{\small[Uri: #1]}}}
\newcommand{\urix}[1]{\textcolor{magenta}{[Uri: \sout{#1}]}}


\newcommand{\toolname}{MutationBERT\xspace}

\definecolor{dkgreen}{rgb}{0,0.5,0}
\definecolor{dkred}{rgb}{0.5,0,0}
\definecolor{gray}{rgb}{0.5,0.5,0.5}
\definecolor{vlgray}{gray}{0.95}
\definecolor{lgray}{gray}{0.7}
\definecolor{bluehighlight}{HTML}{46adb7}
\definecolor{orangehighlight}{HTML}{e3a24d}
\definecolor{redhighlight}{HTML}{b95e8c}

\newcommand{\lstbg}[3][0pt]{{\fboxsep#1\colorbox{#2}{\strut #3}}}
\lstdefinelanguage{JavaDiff}{%
  language     = Java,
  morecomment=[f][\lstbg{red!20}]-,
  morecomment=[f][\lstbg{green!20}]+,
}

\lstdefinestyle{langstyle}{
  basicstyle=\ttfamily\footnotesize,
  keywordstyle=\color{blue},
  commentstyle=\color{dkred},
  stringstyle=\color{dkgreen},
  keepspaces=true,              
  breaklines=true,
  otherkeywords={::=},
  numberstyle=\ttfamily\footnotesize\color{gray},
  stepnumber=1,
  numbersep=8pt,
  backgroundcolor=\color{white},
  tabsize=4,
  showspaces=false,
  showstringspaces=false,
  xleftmargin=.18in,
  captionpos=b,
  escapeinside={(?}{?)},
  frame = single,
}
\lstset{style=langstyle}

\newcommand{\circnum}[1]{\raisebox{.5pt}{\textcircled{\raisebox{-1pt} {#1}}}}

\begin{abstract}

Mutation testing is a powerful technique for
assessing and improving test suite quality that artificially introduces bugs and
checks whether the test suites catch them. However, it is also computationally expensive and 
thus does not scale to large systems and projects. One promising recent approach
to tackling this scalability problem 
uses machine learning to predict whether the tests will detect the synthetic
bugs, without actually running those tests. However, existing predictive mutation testing 
approaches still misclassify 33\% of detection outcomes on a randomly sampled 
set of mutant-test suite pairs. 
We introduce \toolname, an approach for predictive mutation testing that simultaneously encodes the source method mutation and test method,
capturing key \emph{context} in the input representation. Thanks to its higher precision, \toolname saves 33\% 
of the time spent by a prior approach on checking/verifying live
mutants. \toolname, also outperforms the
state-of-the-art in both same project and cross project settings, with meaningful improvements in
precision, recall, and F1 score.  We validate our input representation,
and aggregation approaches for lifting predictions from the test matrix level to
the test suite level, finding similar improvements in performance. \toolname not only enhances the state-of-the-art in 
predictive mutation testing, but also presents practical benefits for real-world applications, both in saving developer time
 and finding hard to detect mutants.

\end{abstract}



\maketitle

\begin{figure*}
  \vspace{2mm}
  \includegraphics[width=1.9\columnwidth]{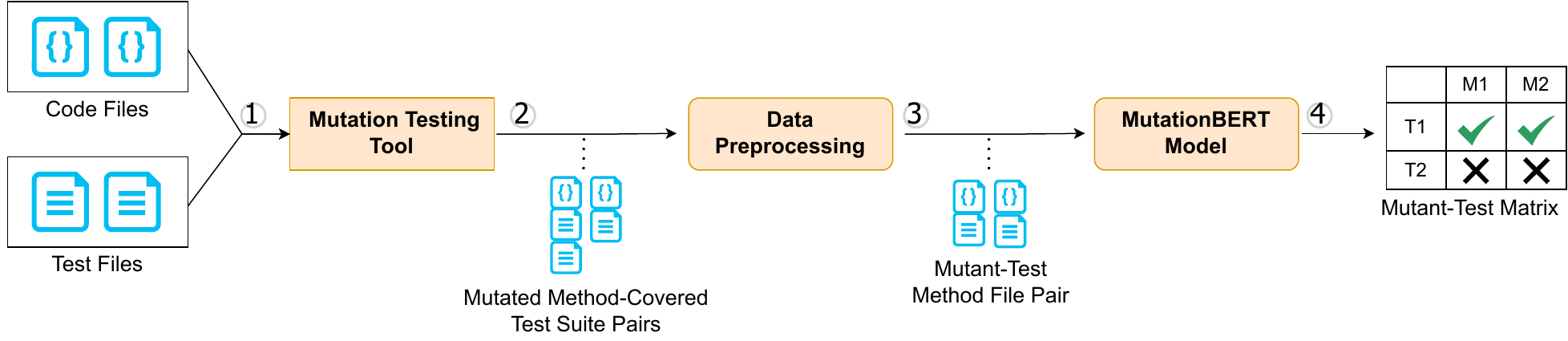}
  \caption{\small An overview of our tool's workflow, including source level mutation tool, preprocessing and model steps. Step \circnum{1} provides source and test files to a mutation testing tool. In Step \circnum{2}, the 
mutation tool uses this information to generate both mutants and their correspondng covering tests. These mutated method-covered test pairs are preprocessed, tokenized, and formatted. In Step \circnum{3}, \toolname takes these inputs to produce 
(Step \circnum{4}) the full mutant-test matrix, the output of our model.}
  \label{fig:workflow-overview}
  \end{figure*}

\section{Introduction}

Mutation testing is a well established technique for evaluating test
suite quality \cite{demillo1978hints, hamlet1977testing,
  jia2010analysis}. Mutation testing works by introducing synthetic bugs based on a fixed set of rules (``mutation operators''), 
ranging from inverting conditional statements to changing unary and binary operators. The test suite is then run on each buggy code copy (also referred to as a ``mutant'' of the original program.  If the test suite fails on a mutant, the mutant is considered ``detected'' (or ``killed''; this is the desired outcome), otherwise the mutant is ``undetected'' (a ``live'' mutant).

Empirically, mutation testing has been shown to improve test suites in ways correlated with real world fault detection~\cite{just2014mutants, papadakis2018mutation}. However, one of its major limitations
is its computational cost: test suites must be run on each mutant, in principle. Large-scale systems commonly have hundreds of thousands of mutants~\cite{icseseip22, GopinathSampleSize}, since mutants scale with size of the codebase and mutation operators considered. 
Myriad approaches, including weak mutation~\cite{HowdenWeak1982}, 
meta mutation~\cite{untch1993mutation}, mutation sampling~\cite{GopinathSampleSize}, and mutant prioritization~\cite{KaufmanFAKAJ2022}, have been proposed to tackle this computational cost. However, they typically require still intractably expensive instrumentation or static and dynamic analyses, and usually rely on some kind of random sampling, compromising their usefulness in practice. 
 Mutation testing has begun to achieve industry
adoption~\cite{BellerFacebookMutation, PetrovicMutationGoogle} at
companies like Meta and Google, leveraging additional heuristics and
idle compute time.  However, current industrial practice is focused on
identifying undetected mutants in newly committed code.  This is, in
essence, the tip of the iceberg; the vast underwater domain of
undetected mutants (and, thus, test weaknesses) in existing code pre-dates the adoption of limited
mutation analysis. Running all mutants on existing large codebases to surface
these problems is still too expensive.

Research on Predictive Mutation Testing\footnote{The first publication~\cite{zhang2016pmt} both named the problem ``Predictive Mutation
  Testing'' and introduced a model/approach to solve it named ``PMT''.  In general in this
  paper, we use ``PMT'' to refer to the problem of predicting
  whether a test/suite will detect a mutant, rather
  than the specific model proposed in that paper.}~\cite{zhang2016pmt, mao2019crossproject, kim2022predictive} takes a different approach to scalable mutation testing, using machine learning to predict whether a mutant will be detected or not \emph{without actually running the tests.} 
The initial PMT work~\cite{zhang2016pmt} empirically demonstrated a correlation between static and dynamic code features and mutant detection, but falls short of practical utility~\cite{AghamohammadiThreatsPMT} in terms of actual F1 or accuracy of the resulting model. 
Seshat~\cite{kim2022predictive} improves on the original PMT model by using ``natural language channels'', including the modified code (pre- and post-mutation), and keywords from the test method and source method name. 
This eliminates the expensive dynamic analyses from the PMT approach
and provides more detailed prediction of which tests detect a mutant
in particular (the mutant-test matrix).
However, although Seshat outperforms the original PMT model, it still suffers from
 significant false postives, with a precision of 0.66 on our test set (Section~\ref{sec:results}), 
 costing valuable developer time.

We observe that there is significant additional \emph{contextual} information embedded in \emph{both} source and test code well beyond simply the mutated line and method names considered in prior work.  By context, we mean both the method surrounding a modified line for a given mutant, as well as the body of the test method, in their entirety.  This intuition is supported by the fact that code and test context are strongly correlated with how useful a mutant is (in terms of whether a mutant is redundant, equivalent, or trivial)~\cite{MutantUtilityContext}.  
In this paper, we build on this insight to enable effective and efficient contextual predictive mutation testing. 

We introduce MutationBERT, a model for predictive mutation testing that takes as input the mutated source method and corresponding test method. MutationBERT learns the relationship between them to predict whether the test will fail on that modified method. 
To this end, we introduce a novel input representation that encodes each mutation as a token level diff applied to a source method, followed by the corresponding test.
We then use a pretrained transformer~\cite{vaswani2017attention} architecture to encode source and test methods, and further finetune it for our task. 

A transformer maps a sequence of tokens to a contextual embedding that can subsequently be finetuned to downstream tasks. 
Transformers have been shown to be highly effective across a wide range of software engineering tasks, ranging from code completion to 
merge conflict resolution~\cite{wang-etal-2021-codet5, svyatkovskiy2021mergebert, ahmad-etal-2021-unified, feng-etal-2020-codebert}. Their highly parallel architecture means that inference time is low, as
compared to RNNs used in prior work in predictive mutation testing~\cite{kim2022predictive}. To our knowledge, our work is the first to apply this recent advancement to this domain.
As implied by the name, MutationBERT builds on recent advancements in pretrained code models by finetuning CodeBERT~\cite{feng-etal-2020-codebert} for mutation testing. Due to having seen so much code, pretrained models have a better representation and understanding of source code syntax and
semantics, and thus are better equipped for tackling source-intensive tasks such as mutation testing.

Like Seshat, MutationBERT requires no computationally expensive static or dynamic analysis, nor instrumentation, as MutationBERT operates
entirely on source text. MutationBERT can also generate the full mutant-test matrix.
Generating the full matrix
is essential for many applications of mutation testing.  For example, if a
mutant is predicted to be detected by only a very small number of
tests, the prediction can be confirmed by running just those tests.
Mutants predicted to be undetected can similarly be checked by running
the tests considered most likely (though still unlikely) to detect
them.  Importantly in practice, a developer who wants to add testing to cover an
undetected mutant will certainly want to know which existing tests
would be most likely to detect the mutant, since often the way to fix
such a problem is to strengthen the oracle or extend the behavior of
an existing test. 

To summarize, our core contributions are as follows:
\begin{itemize}
    \item We perform an extensive empirical evaluation of predictive mutation testing tools, including our approach, \toolname, and Seshat. We measure both inference time and the runtime cost difference between both tools. 
    We consider the tradeoff between precision and recall here and discuss its impact on the end user, finding that our higher precision saves 33\% of the total time spent checking mutants. We also investigate 
    the performance of both tools in detecting non-trivial mutants, finding \toolname has a 30\% improvement in accuracy over Seshat.
    \item We introduce \toolname, the first (to our knowledge) predictive mutation testing model to incorporate source and test code context. \toolname can predict entire mutant-test matrices along with whether mutants
    are detected or undetected by test suites. We find that \toolname has a 8\% improvement in F1 score when predicting test matrices, and over a 12\% improvement 
    in F1 score over Seshat (the state-of-the-art baseline) when predicting whether a mutant is detected or undetected by the entire test suite. While recall remains relatively stable, precision improves
    25\% between Seshat and \toolname, meaning that mutants labeled as undetected by \toolname are much less likely to be false postives.
    \item We perform an extensive analysis of the design decisions, including an examination of alternative
    input representations that leverage both source and test method context. We find that token-level diff is the most effective input representation for mutant prediction.
\end{itemize}

We release our dataset, source code, and model checkpoints at \url{https://doi.org/10.5281/zenodo.7600371}, including detailed instructions on how to reproduce all of our results and use our model. We hope that this will
enable the community to deploy our model and further build upon our work.


\section{Contextual Predictive Mutation Testing}
\begin{figure*}
  \centering
  \begin{subfigure}{0.4\linewidth}
    \centering
\begin{lstlisting}[language=JavaDiff,numbers=left]
public RegularTimePeriod next() {
  Hour result;
-  if (this.hour != LAST_HOUR_IN_DAY) {
+  if (this.hour > LAST_HOUR_IN_DAY) {
    result = new Hour(this.hour + 1, this.day);
  }
  ...
}

public void testNext() {
  Hour h = new Hour(1, 12, 23, 2000);
  h = (Hour) h.next();
  assertEquals(2000, h.getYear());
  ...
}
\end{lstlisting}    
    \caption{\small Motivating example}
    \label{fig:motivating-example}
  \end{subfigure}%
  \hfill
  \begin{subfigure}{0.45\linewidth}
    \centering
\begin{lstlisting}[language=Java,numbers=left]
(?\colorbox{brown!50}{<CLS>}?) 
public RegularTimePeriod next() {
  Hour result;
  if (this.hour (?\colorbox{brown!50}{<BEFORE>}?) != (?\colorbox{brown!50}{<AFTER>}?) > (?\colorbox{brown!50}{<ENDDIFF>}?) 
    LAST_HOUR_IN_DAY) {
    result = new Hour(this.hour + 1, this.day);
  }
  ...
}
(?\colorbox{brown!50}{<SEP>}?)
public void testNext() {
  Hour h = new Hour(1, 12, 23, 2000);
  h = (Hour) h.next();
  assertEquals(2000, h.getYear());
  ...
}
\end{lstlisting}
    \caption{\small Model encoding of example}
    \label{fig:model-encoding}
  \end{subfigure}
  \caption{\small A snippet of code from the popular JFreeChart Java project, where a mutation changing \texttt{!=} to \texttt{>} is applied (\Cref{fig:motivating-example}). The provided 
  test fails to detect this mutant. \Cref{fig:model-encoding} shows how we encode this mutant in our approach. Newly added special tokens are marked in \colorbox{brown!50}{brown}.
  }
  \label{fig:example}
  \end{figure*}

Figure~\ref{fig:workflow-overview} overviews the MutationBERT workflow. 
Our workflow takes a project and test
suite as input, and uses a given source-level mutation testing tool (step
\circnum{1}, Section~\ref{sec:mutation-testing}) to generate a set of mutants and tests that cover them (step \circnum{2}). 
Most mutation testing tools provide coverage out of the box, as a way to
  prune uncovered mutants, which will always be undetected. We encode the method/test pairs in an
input representation (step \circnum{3}, Section~\ref{sec:ir}), to be passed as
input to our trained model (step \circnum{4}, Section~\ref{sec:model}).  The
model predicts whether the test will detect or fail to detect the mutant (step
\circnum{5}). 
Over all mutant-test pairs, these predictions comprise the mutant-test 
matrix for the program.  This output can be optionally post-processed to
aggregate predictions across the whole test suite.  This produces for the user a
set of mutants 
likely undetected by the test suite; these can be inspected directly, or ranked
by existing mutant prioritization algorithms~\cite{KaufmanFAKAJ2022, PetrovicMutationGoogle, BellerFacebookMutation}. As the developer adds tests, more
interesting mutants are identified, leading to better test suites over time.

As an illustrative example, consider
\Cref{fig:motivating-example}, which shows a (simplified) code and test
snippet from  JFreeChart.\footnote{\url{https://github.com/jfree/jfreechart}} The
\texttt{next()} method returns the next hour for a given
\texttt{RegularTimePeriod}. The \texttt{testNext} method checks that it works
correctly for 23:00 on December 1st, 2000. Although this test method may look
comprehensive, note that it does not fail if we change the \texttt{!=} operator to \texttt{>}
on line 3. A better test suite would include another method that
includes a time that is not the last hour of a day, which would correctly fail
on the mutated code. We will refer to this example throughout subsequent
sections to clarify our contribution. 

\subsection{(Predictive) Mutation testing}
\label{sec:mutation-testing}

Mutation testing \cite{demillo1978hints} is the process of
synthetically introducing faults into programs and measuring the effectiveness
of tests in catching them. A set of program transformations, known as ``mutation
operators'' take regular code and create buggy copies of it. These operators
vary~\cite{just2014Major, universalMutator, colesPIT2016}, but some common
operators include negating conditions (\texttt{if (a)} to \texttt{if (!a)}),
replacing arithmetic operators (\texttt{a + b} to \texttt{a - b}), replacing
relational operators (\texttt{a < b} to \texttt{a > b}), and flipping
conditionals (\texttt{a == b} to \texttt{a || b}). Each time one of these rules
is applied to a program, a new \emph{mutant} is created, each differing only slightly
from the original program. The change in \Cref{fig:motivating-example}
creates one such mutant for the \texttt{next()} method.

Test adequacy is measured by running the entire test
suite on each mutant; the goal is a test suite that detects all mutants,
increasing confidence that the suite would detect unintentional bugs as well.
The test suite corresponding to the single test \texttt{testNext()} in
\Cref{fig:motivating-example} does not detect the mutant; presenting this mutant
to a developer would ideally motivate them to create the necessary additional
tests.  
Mutation score, or the ratio of detected mutants to total mutants, provides a
rough measure of test adequacy, outperforming code coverage in terms of
correlation with real-world fault detection~\cite{just2014mutants,
  papadakis2018mutation}. Mutation testing has seen some industry
adoption~\cite{PetrovicMutationGoogle,BellerFacebookMutation}.  Prominent recent
uses at Facebook and Google apply it only to changed code at commit-time, which
still requires large amounts of idle compute~\cite{petrovicIndustrialChallenges2018}
because of the massive computational expense of running it over an entire codebase.
Tackling this scalability problem
\cite{BokaeiChallenges2019} is the core motivation of our work.

Our approach is parametric with respect to existing source-level mutation
testing tool and can integrate with existing approaches like
Major \cite{just2014Major} and universalmutator \cite{universalMutator}. For our evaluation we use
a set of mutants collected by Major \cite{just2014Major} on the Defects4J 2.0
dataset provided by \citet{kim2022predictive} with the Seshat experiments.

Techniques for Predictive mutation testing~\cite{zhang2016pmt, mao2019crossproject,
  kim2022predictive} use machine learning to predict whether a test or a test
suite will detect a mutant without actually running those tests.  We provide
more detailed comparison in Section~\ref{sec:related}.  For the purposes of
understanding our technique, however, note that one limitation of the first
ML-based approach for mutation testing prediction~\cite{zhang2016pmt} is that
its performance degrades significantly when it is not trained/evaluated on mutants that
are not covered (executed) by any of the tests in the test
suite~\cite{AghamohammadiThreatsPMT}.  Uncovered mutants are trivially
undetected by a test suite, since a test cannot fail due to a bug on a line it
does not execute.  They are thus not interesting for the task of predictive
mutation testing.  As a result, our approach follows precedent set in subsequent
work~\cite{kim2022predictive} and excludes
uncovered mutants from the prediction task. 

\subsection{Input representation}
\label{sec:ir}

Our goal is to train a model that predicts whether a given test will detect a
given mutant.  Concretely, a mutant is a typically small modification to a
typically much larger code file.  Prior efforts to represent code changes for
the purpose of ML, fall into three main categories: defining a set of features related to the modification~\cite{kim2022predictive, zhang2016pmt} 
representing the modification with a graph~\citep{maGraphCode2Vec2022, YasunagaGraphBased2020, WenhanCodeClone2020} or representing the ``before'' and ``after'' of the modification with 
multiple embeddings~\cite{svyatkovskiy2021mergebert}. 

For earlier PMT models \cite{kim2022predictive, zhang2016pmt} 
that did not use pretrained transformers, defining a set of features and aggregating them into a single vector made sense. However, to leverage the gains from using a 
pretrained model like CodeBERT~\cite{feng-etal-2020-codebert}, we need to represent our inputs in the same way as the pretrained model, making the feature-based approach unviable.
Following best practices in pretrained transformers, we use the same input embeddings for encoding the mutated code and the tests.

Thus, we represent each mutant-test pair as a token level diff to \toolname, using the special tokens \texttt{<BEFORE>}, \texttt{<AFTER>} and \texttt{<ENDDIFF>}. 
For example, if the line \texttt{...if a == b:...} is changed to \texttt{...if a != b:...}, we encode it in the following
manner: \texttt{...if a <BEFORE> == <AFTER> != <ENDDIFF> b:...}. This encode diffs compactly, while preserving original code structure.

\Cref{fig:model-encoding} shows how our model encodes the motivating example. We provide the model with the source method encoded as a token-level diff, followed by the test method. 
Our model then outputs whether such a mutant is detected or undetected. We follow CodeBERT~\cite{feng-etal-2020-codebert} in their use of special tokens \texttt{<CLS>} and \texttt{<SEP>}.
CodeBERT uses \texttt{<CLS>} and \texttt{<SEP>} to denote code and natural language input, using \texttt{<CLS>} token for downstream classification tasks (we discuss this in more
detail in Section \ref{sec:model}). Similarly, we separate code and test with the special \texttt{<SEP>} token. We take the hidden representation of the \texttt{<CLS>} token as the vector which we train the 
model to classify whether this mutant is detected or not.

\subsection{Model}
\label{sec:model}

Our model can predict either the entire mutant-test matrix for a project, or
whether a single mutant is detected by an entire test suite.
Our model is a pre-trained CodeBERT model fine-tuned to the mutation
testing task, with a novel input representation.
CodeBERT \cite{feng-etal-2020-codebert} is a pretrained model that leverages the transformer architecture \cite{vaswani2017attention}. 
It was trained to predict \emph{masked} tokens (code or natural language tokens replaced with \texttt{<MASK>}) for both source code and natural language.
CodeBERT uses special \texttt{<CLS>} and \texttt{<SEP>} tokens to denote code
and natural language, using the \texttt{<CLS>} token for classification in
downstream tasks. 
CodeBERT was pretrained on a corpus of 6.4 million functions across seven
different programming languages; large pretrained models like CodeBERT are applicable to a variety of downstream tasks ranging from code completion \cite{feng-etal-2020-codebert}, to merge conflict resolution \cite{svyatkovskiy2021mergebert}, 
and code summarization \cite{ahmad-etal-2021-unified}. To the best of our knowledge, we are the first to leverage pretrained models for the task of predictive mutation testing.

We formulate mutation analysis as a binary classification task to CodeBERT. We provide CodeBERT with both the source
method encoded as a token level diff and the test method (Section~\ref{sec:ir}).
After feeding the input to CodeBERT, we pass the encoding of the \texttt{<CLS>}
token through a linear layer, 
which is then used to make the final classification. The model is called for
each mutant-test pair to construct the entire mutant-test matrix.  

We use the probability output of the model to aggregate predictions across each mutant's set of covered tests, and consider a mutant to be ``detected'' if the confidence of the model on at least \emph{one} of the tests is greater than 0.25:

\begin{equation}
  \mathtt{pred}_{M,T} = \begin{cases}
    \text{``detected''} & \left(max_{t \in T}MutationBERT\left(M, t\right)\right) > 0.25 \\
    \text{``undetected''} & \text{otherwise}
  \end{cases}
\end{equation} 

where $M$ corresponds to the mutant and $T$ corresponds to the set of tests that cover the mutant. We chose 0.25 as our confidence threshold, as it was able 
to reduce the number of false positives when evaluated on our validation dataset, with a precision of 0.76, while not reducing the overall \emph{F1} score of 0.80. 

\section{Experimental Setup}

We compare \toolname with Seshat~\citep{kim2022predictive}, the current
state-of-the-art model for PMT, using the dataset from that paper.
We ask the following research questions:

\vspace{1ex}
\noindent\textbf{RQ1: Effectiveness: How well does \toolname perform in a \emph{same project} setting?}
In a \emph{same project} setting, a PMT model is trained on previous
versions of a project, and then used to predict test matrices, unkilled mutants,
or mutation scores for subsequent versions.  We compare \toolname to Seshat on a
within-project task, evaluating the models' correctness when predicting
test-mutant matrices and over the test suite- level
aggregation. 

\vspace{1ex}
\noindent \textbf{RQ2: Generality: How well does \toolname perform in a \emph{cross project} setting?}
In a \emph{cross project} setting, a PMT model is trained using data from one
project and then used to predict test-mutant behavior for a different project.
This is much more difficult than the same project setting, but could be
especially applicable when starting a new project, for example.  We compare
\toolname to Seshat on the cross-project task using the same metrics as the
\emph{same project} task. 

\vspace{1ex}
\noindent\textbf{RQ3: Design Decisions: How do different input representations and aggregation approaches affect our final model?}
We analyze and compare several input representations as well as aggregation
approaches to validate the design decisions underlying \toolname. 

\vspace{1ex}
\noindent\textbf{RQ4: Qualitative Analysis: What are causes of \toolname mispredictions?}
We manually examine 100 cases where our model misclassifies a mutant as detected
or undetected to identify common reasons for failures and better understand
limitations.

\vspace{1ex}
\noindent\textbf{RQ5: Efficiency: How efficient is \toolname compared to prior work, and
  regular mutation testing?}
We address how \toolname compares to Seshat, and characterize the performance
improvement it provides over regular mutation testing. 

\vspace{1ex}
\noindent\textbf{RQ6: Mutant Importance: How effective is \toolname at
  predicting difficult-to-detect mutants (that fail only a small number of tests)?}

We address how \toolname compares to Seshat with regards to how many tests
detect a given mutant, a proxy for mutant difficulty. 

\subsection{Baseline}

We compare against the Seshat baseline \cite{kim2022predictive}. Seshat is a
state-of-the-art model for mutation testing, which has been shown to outperform
PMT \cite{zhang2016pmt} by 0.14 to 0.45 \emph{F1} score depending on project.
Similar to our model, Seshat has no overhead in static or dynamic analysis,
operating entirely on source level features, unlike the prior model PMT, which
requires both static 
and dynamic analysis to run. However, unlike our model, Seshat operates over a
set of features: the source method name, the test method name, the mutated line
before and after, and a one-hot encoding of the mutation operator. 
Seshat first encodes the source and test method names with a bidirectional GRU.
It then concatinates the resulting embeddings with a one-hot encoding of the mutation 
operator to classify the mutant as detected or undetected by the test.    

Like our model, Seshat outputs a confidence score for each mutant-test pair,
which we aggregate to predict whether the mutant is detected or not by the
entire test suite. 
We aggregate Seshat's predictions across each mutant's set of covered tests by
comparing confidence to a threshold.  We set this threshold to 0.10, which in
our experiments produced the highest \emph{F1} score for Seshat in validation
(Seshat does not mention a a threshold in their paper, so we perform the same
optimization as we did for \toolname). 
We thus aggregate as follows:

\begin{equation}
    \mathtt{pred}_{M,T} = \begin{cases}
    \text{``detected''} & \left(max_{t \in T}Seshat\left(M, t\right)\right) > 0.10 \\
    \text{``undetected''} & \text{otherwise}
  \end{cases}
\end{equation}
where \texttt{M} corresponds to the mutant and \texttt{T} corresponds to the set of tests that cover the mutant.

\subsection{Dataset}

\begin{table}
  \caption{\small Our dataset comprising of 6 Defects4J 2.0 projects.\label{tab:data}}
\begin{tabular}{l|lrr}
\toprule
\bf Project                  & \bf Date            & \bf LOC   & \bf \#tests        \\
\midrule
commons-lang                 & 2013-07-26           &  21,788 &  2,291 \\
jfreechart                   & 2010-02-09           &  96,382 &  2,193 \\
gson                         & 2017-05-31           &  7,826 &  1,029 \\
commons-cli                  & 2010-06-17           &  2,497 &  354 \\
jackson-core                  & 2019-01-06           &  25,218 &  573 \\
commons-csv                  & 2017-12-11           &  1,619 &  290 \\
\bottomrule
\end{tabular}
\end{table}

  \begin{table}
    \caption{\small Tests, mutants and mutant-test pairs (pairs) for both same project and cross project settings, across training (train), validation (val), and test (test) sets. Note that mutant-test pairs only include tests that cover a given mutation.}
    \begin{center}
        \begin{tabular}{ll|rrr}
            \toprule
  &  \textbf{Split} & \textbf{\#tests} & \textbf{\#mutants} & \textbf{\#pairs} \\
  
            \midrule
  \multirow{3}{*}{Same Project} &          train & 6,124 &  68,702 & 1,522,924 \\
            &  val   & 5,644 & 8,688 & 197,527    \\
            &  test  & 5,637 & 8,648 & 195,140   \\
  \midrule \midrule 
  \multirow{3}{*}{Cross Project} &      train & 4,725 & 79,128 & 1,460,344 \\
            &  val   & 1,171 & 5,427 & 402,296\\
            &  test  & 261 & 1,040 & 42,687 \\
            \bottomrule
            \end{tabular}
            \label{tab:dataset-splits}
    \end{center}
    \end{table}  

We reuse the dataset released with the Seshat
experiments~\cite{kim2022predictive}. This dataset consists of a full mutation
analysis in Major \cite{just2014Major} of six large scale Java projects, with
extensive testing, across multiple versions, taken from Defects4J v2.0.0 (statistics shown in 
\Cref{tab:data}).
This dataset considers only mutants that are actually covered by some test,
since uncovered mutants cannot be detected by a given test suite (and
can be discarded with a simple coverage heuristic).

Note that the Seshat evaluation~\cite{kim2022predictive} analyzed the
cross-version setting in detail, training models on previous versions of
programs to predict matrices for subsequent versions. The models remain
effective across versions many years apart. This is likely a function of the
fact that code (and mutation behavior) is quite stable over time, as shown in
the dataset description in \citet{kim2022predictive}.

Thus, in the interest of space and computational effort, we restrict our
attention to single versions per project for all RQs. We select the latest
versions of the six projects in Defects4J 2.0 and perform a 80-10-10 split
between train, validation and test sets. In the same project setting, we split
by mutant-test suite pair. This is in contrast to the prior evaluation, that is,
mutant-test pairs from the \emph{same} test suite must be part of the same
subset. Practically, our envisioned application does not include a situation
where a PMT model could be trained on data corresponding to whether half the
tests in a given test suite detect a given mutant, and then used to predict the
behavior of the other half. This explains why we reran Seshat (and why our
numbers may not match those in the original paper).
For the cross project setting, we split by project, where each project consists
of a set of mutant-test suite pairs. 
We use the exact same splits for our model and for Seshat. \Cref{tab:dataset-splits} shows statistics about our 
same project and cross project splits.


\subsection{Preprocessing and Training} 

We use the pretrained RoBERTa tokenizer (BPE tokenizer \cite{sennrich-etal-2016-neural}) with vocabulary size of 50,000 tokens for all programming languages that is provided with CodeBERT.
We finetune CodeBERT with context window size of 1024 tokens. Thus we only show \toolname the first 1024 tokens of the code and test if their combined length is longer than 1024 tokens. Such cases
account for 14.6\% of all mutant test pairs. 

We follow the same steps that \citet{kim2022predictive} took to train Seshat. We train Seshat for 10 epochs, with a batch size of 512, and learning rate of \texttt{3e-3}. We train \toolname 
for eight epochs with learning rate of \texttt{1e-5} and batch size of 64. We use a weighted loss function according to the distribution of detected and undetected mutant-test pairs.
We use a linear warmup to 1000 steps, followed by a cosine annealing decay, in accordance with best practices for fine tuning transformers \citep{PopelTrainingTips}. 
Both models' loss functions converge using these settings. We fine-tuned our model on a Nvidia GeForce RTX 3080 for one week for a total of 115k steps. 

\subsection{Metrics and Settings}
\label{sec:metrics}

One way to use models for predictive mutation testing is to compute mutant-test
matrices, which predict, for each mutant, whether each test passes or fails.  
In general, most tests pass on most mutants.  That is, a test detecting a mutant
is the minority class.  In this setting, model \emph{precision} refers to how accurately 
mutants are identified as detected, while \emph{recall} refers to the proportion of detected mutants labeled correctly. In the mutant-test matrix
setting 72\% of mutant-test pairs are undetected. We care that our model is able to accurately 
predict the remaining 28\% of detected mutants; the goal is to
identify the few tests that detect each mutant. 

Another way to use these models is to predict whether an entire test suite
detects a particular mutant.  Here, the majority class is detected mutants; 61\% of 
mutants are detected. The core goal here is to accurately identify the undetected mutants, to guide
developers to improve test suites. Therefore, we define precision and recall differently than in the the mutant-test
matrix setting.  In the test suite setting, model \emph{precision} refers to how accurately 
mutants are identified as \emph{undetected}, while recall refers to the proportion of \emph{undetected} mutants
that are classified correctly. \emph{Precision} is thus important in understanding the potential cost
of a PMT model in terms of time needed to either actual run the test suite to
confirm its predictions, or time wasted by a developer inspecting an ultimately
uninteresting mutant. \emph{Recall} is also important to overall model usefulness: 
if a model misses a large number of undetected mutants, key gaps in test suite quality could remain.  

We report precision, recall and F1 score (which balances the two) for
all models in the first three research questions. For RQ1 (same project) and RQ2
(cross project), we evaluate performance both on the base test set (195,140
mutant-test pairs). For efficacy of prediction over the entire test suite, we evaluate \toolname on the same
dataset, aggregated at the test suite level (8648 test suites).

For RQ3, we evaluate different aggregation thresholds and input representation
choices on the validation set consisting of 120,710 mutants, again reporting
precision, recall, and F1 scores; we evaluate both mutant-test predictions and
mutant-test suite predictions. Due to compute constaints associated with a larger context window, 
we use the smaller 512 token context window of CodeBERT and respective dataset 
to evaluate different thresholds and input representations.

For RQ4, to ensure a representative sample of misclassifications, we randomly
select 100 examples where our model misclassifies a mutant as being detected or
undetected. We manually examine each example and try to understand the cause of
the misprediction. Finally, we bucket these mispredictions in a series of
categories and discuss these in detail. We do this to inform a general assay of
the limitations of our technique; we do not make strong claims about the
generalizability of this qualitative assessment.

For RQ5, we run 1000 iterations of Seshat and \toolname, with a batch size of one, on a workstation with an
Nvidia GeForce RTX 3080 GPU, with 100 warmup iterations. We report the average
time taken over these 1000 iterations as the inference time for each model.
To compute comparative time and speedups against regular mutation testing, we 
use numbers from previous work~\cite{kim2022predictive} in conjunction with our 
inference time numbers.

For RQ6, we report accuracy of Seshat and \toolname with respect to percentage of tests
that kill a mutant. The goal is to measure whether \toolname is only correctly classifying "easy" to detect
or "trivial" mutants where the majority of tests detect the given mutant or whether \toolname is capable
of correctly classifying mutants that are more difficult to detect.

\section{Results and Analysis}
\label{sec:results}

  \begin{table*}
    \caption{\small Comparison between Seshat and \toolname on both same project and cross project settings in terms of precision, recall and F1 score. In both same project and cross project settings, \toolname outperforms
    Seshat across all metrics, with an \emph{F1} score difference of 12\% on the same project setting and \emph{F1} score difference of 28\% on the cross project setting.}
    \begin{center}
        \begin{tabular}{l|l|rrr|rrr}
            \toprule
             \multirow{2}{*}{\textbf{Setting}} & \multirow{2}{*}{\textbf{Model}}  & \multicolumn{3}{|c|}{\textbf{Mutant-Test Matrix}} & \multicolumn{3}{|c}{\textbf{Test Suite}}\\ \cline{3-8}
             & & \textbf{Precision} & \textbf{Recall} & \textbf{F1} & \textbf{Precision} & \textbf{Recall} & \textbf{F1}  \\
  
            \midrule
            \multirow{2}{*}{Same Project} & Seshat  & 0.66 &  0.68 & 0.67 & 0.56 & \textbf{0.82} & 0.67 \\
            & \textbf{\toolname} & \textbf{0.72} & \textbf{0.77} & \textbf{0.75} & \textbf{0.81} & 0.78 &  \textbf{0.79} \\
            \midrule
            \multirow{2}{*}{Cross Project} & Seshat & 0.58 & 0.29 & 0.38 & 0.24 & 0.39 & 0.30 \\
            & \textbf{\toolname} & \textbf{0.68} & \textbf{0.37} & \textbf{0.48} & \textbf{0.52} & \textbf{0.65} & \textbf{0.58} \\
            \bottomrule
            \end{tabular}
            \label{tab:results-test-set}
    \end{center}
    
    \end{table*}

We report results for all five RQs, and discuss their implications.

\subsection{RQ1: Same Project Performance}

\Cref{tab:results-test-set} shows the results of \toolname and Seshat on the
test set for the \emph{same project} setting.  The center columns show results
in predicting whether a test will detect a particular mutant, relevant to
constructing the overall mutant-test matrix.  \toolname outperforms Seshat
across all metrics: \toolname's \emph{F1} score is 0.75, compared to Seshat's
0.67. Interestingly, \toolname and Seshat
have similar precision (0.66 for Seshat vs 0.72 for \toolname); the models
report similar numbers of false positives (cases where the models misclassify a test as detecting a mutant).  
However, \toolname has higher recall
(0.77, versus 0.68), meaning that \toolname is more likely to correctly identify
cases where a test detects a mutant. 

When the predictions are aggregated into test suite level predictions (right-hand columns), 
recall that undetected mutants are the minority class, flipping the meaning of
precision and recall (Section~\ref{sec:metrics}).  Seshat and \toolname both find similar
numbers of undetected mutants, but
\toolname has much higher precision, 0.81, compared to Seshat's 0.56.  False positives are costly, as
they cost developers valuable time examining mutants that are in reality detected by their test suite.

Another way of viewing these results is in terms of the difference between the
mutation score estimated by a predictive mutation model, and the actual
mutation score.  Recall that mutation score is the true ratio of detected
mutants to total mutants; empirically, mutation score provides a
better measure of test adequacy than code coverage~\cite{just2014mutants,
  papadakis2018mutation} and thus is useful (albeit usually expensive) to compute. 
The gold mutation score (true mutation score) on our test set is 0.59.  Seshat
estimates a mutation score of 0.40 over the entire dataset, an error of 0.19.
\toolname computes a mutation score of 0.61, a difference of only 0.02 from the
true answer.  \toolname thus has much lower error in estimating mutation score
on this dataset as compared to Seshat. 


\subsection{RQ2: Cross Project Performance}

\Cref{tab:results-test-set} also shows the \emph{cross project} setting (bottom
rows), where a model is trained on one set of projects and evaluated on another.
Again, \toolname outperforms Seshat  (0.68 precision and 0.37 recall for \toolname and 
0.58 precision and 0.29 recall for Seshat). That said, in the mutant-test predictions, both
precision and recall drop significantly for both approaches; this suggests that
training data containing project-specific vocabulary and methods contribute
substantially to the same project performance. This is consistent with other
results showing that projects have distinct vocabulary and style, 
making cross project prediction difficult for many tasks~\cite{hellendoorn2017deep, Ahmed2022FewshotTL}. 
Precision continues to be quite a bit higher than recall in the cross project
setting, for both models. 

At the test suite level, we find that \toolname outperforms Seshat on all
metrics. Precision is very low for both tools; Seshat and \toolname both misclassify
a significant proportion of undetected mutants, however \toolname has a significantly 
higher precision. 
Recall is also low in the cross project setting, at 0.39 for Seshat
and 0.65 for \toolname.  
However, this indicates that in a cross project setting \toolname is capable of finding more 
undetected mutants than Seshat. 

On the cross project test set, the gold mutation score is 0.77. Seshat differs
from this value significantly, with a mutation score of 0.63 (error of 0.14). \toolname 
is much closer, predicting a mutation score of
0.72 (error of 0.05).


\subsection{RQ3: Input Representations and Aggregation Approaches}

We proposed a new input representation for the mutation prediction
problem.  Here, we describe several alternatives that we then experimentally
evaluate.  We also describe alternative aggregation approaches. Then, we 
evaluate these alternatives (all on the validation
set) to motivate the input representation and aggregation approaches 
in our final model. 

\begin{figure*}
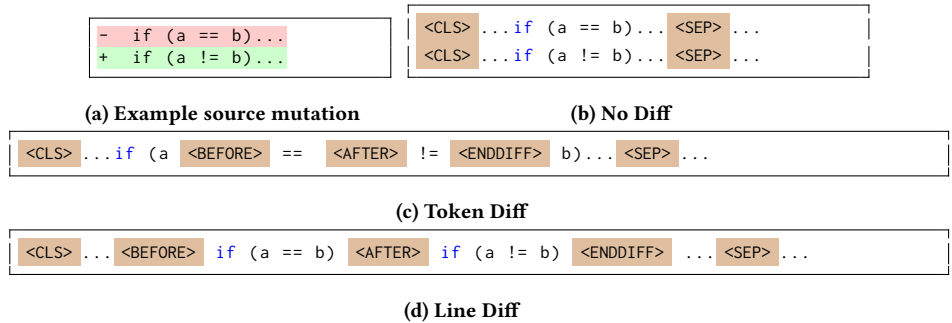

  \centering
  \begin{subfigure}{0.5\columnwidth}
    \centering
\begin{lstlisting}[language=JavaDiff]
-  if (a == b)...
+  if (a != b)...
\end{lstlisting}    
    \caption{\small Example source mutation}
    \label{fig:source-mutation}
  \end{subfigure}%
  \begin{subfigure}{0.75\columnwidth}
    \centering
\begin{lstlisting}[language=Java]
(?\colorbox{brown!50}{<CLS>}?)...if (a == b)...(?\colorbox{brown!50}{<SEP>}?)...
(?\colorbox{brown!50}{<CLS>}?)...if (a != b)...(?\colorbox{brown!50}{<SEP>}?)...
\end{lstlisting}
    \caption{\small No Diff}
    \label{fig:no-diff}
  \end{subfigure}
  \begin{subfigure}{1.5\columnwidth}
    \centering
\begin{lstlisting}[language=Java]
(?\colorbox{brown!50}{<CLS>}?)...if (a (?\colorbox{brown!50}{<BEFORE>}?) ==  (?\colorbox{brown!50}{<AFTER>}?) != (?\colorbox{brown!50}{<ENDDIFF>}?) b)...(?\colorbox{brown!50}{<SEP>}?)...
\end{lstlisting}
    \caption{\small Token Diff}
    \label{fig:token-ir}
  \end{subfigure}
  \begin{subfigure}{1.5\columnwidth}
    \centering
\begin{lstlisting}[language=Java]
(?\colorbox{brown!50}{<CLS>}?)...(?\colorbox{brown!50}{<BEFORE>}?) if (a == b) (?\colorbox{brown!50}{<AFTER>}?) if (a != b) (?\colorbox{brown!50}{<ENDDIFF>}?) ...(?\colorbox{brown!50}{<SEP>}?)...
\end{lstlisting}
    \caption{\small Line Diff}
    \label{fig:line-ir}
  \end{subfigure}
  \caption{\small Input representations for encoding mutations applied to source code. Each subfigure shows a different input representation on the same example of changing \texttt{==} to \texttt{!=}.
  Token diff and line diff were the best performing input representations and we chose to use token diff as the final input representation in \toolname.}
  \label{fig:input-representations}
  \end{figure*}

\subsubsection{Input Representations}
We outline various input representations that incorporate source and test context for our model. For all input representations, we separate method code and test code with a
<CLS> token, which we use for classification.

\noindent \textbf{No Diff (Binary Task):} Our simplest approach is to directly apply the mutation and feed the model both the mutated version of the code and unmutated version of the code. For example,
when changing \texttt{==} to \texttt{!=} in \texttt{...if a == b:...} we feed the model both \texttt{...if a == b:...} and \texttt{...if a != b:...} (Figure \ref{fig:no-diff}). 

Since we have likelihood scores for both the mutated and unmutated versions of the code, we try two modes of evaluation. Our first mode feeds the model the mutated code, and takes its prediction. Our second mode feeds the model both
the mutated code and unmutated code and obtains its probability of being detected. Then it subtracts these two probabilities from each other (since we know the first datapoint is always undetected),
and compares this difference against a dynamically set threshold. We try all thresholds between 0.01 and 0.99 in increments of 0.01 on the validation set, and select the best performing threshold.

\noindent \textbf{Token Level Diff:} We represent each mutation as a token level diff. For example if a line \texttt{...if a == b:...} is changed to \texttt{...if a != b:...}, we encode it in the following
manner: \texttt{...if a <BEFORE> == <AFTER> != <ENDDIFF> b:...} (Figure \ref{fig:token-ir}). This allows for the most compact footprint in encoding the diffs, allowing our model to learn 
how certain diffs coupled with the surrounding code and test are correlated with a mutant being detected or not detected.

\noindent \textbf{Line Level Diff:} For line level diffs, we represent diffs in terms of change to source lines. This input representation is similar to token level diff. In our example, we encode
the mutation as \texttt{...<BEFORE> if a == b: <AFTER> if a != b: <ENDDIFF> ...} (Figure \ref{fig:line-ir}). We hypothesize that this might perform better than token diff, as CodeBERT was 
pretrained for tasks such as next line prediction.

\subsubsection{Aggregation Approaches}

We outline aggregation approaches that we tried for our test matrix model. Practically, this aggregation holds value, as undetected mutants
(mutants not detected by the entire test suite) are ones of interest to developers, as they indicate testing inadequacy. Specifically, in order to use such a model,
aggregate predictions need to be accurate, otherwise undetected mutants will be identified incorrectly.

\noindent \textbf{Threshold Aggregation:} We aggregate the predictions of both predictive mutation testing models by using various probability thresholds (0.1, 0.25, 0.5, 0.75 and 0.9). 
Specifically, we only label a test as detecting a mutant if the model predicts the test detects the mutant with probability above the defined threshold. We vary thresholds to 
observe their effect on precision, recall, and F1 score. 

\noindent \textbf{Learned Aggregation:} We also tried learning an aggregation based off of the embeddings of the <CLS> token after CodeBERT encoding. We use a transformer with three layers
 to take these embeddings and aggregate them. We then use a linear layer to classify based off of this learned aggregate embedding whether the test suite detects or fails to detect
 the mutant. We evaluate this learned aggregation both using a weighted loss function (according to the data distribution) and using a normal loss function.

\subsubsection{Experimental Results}

\begin{table}
\caption{\small Precision, recall and \emph{F1} scores of all models at predicting the mutant-test matrix on the validation set. Token diff and line diff are the best performing models, with an F1 score of 0.78.}  
\centering
\begin{tabular}{l|rrr}
\toprule
\bf Model                  & \bf Precision            & \bf Recall   & \bf F1        \\
\midrule
Seshat                & 0.73                     & 0.75   & 0.74               \\
\bf Token Diff          & \bf 0.79                     & \bf 0.77     & \bf 0.78               \\
\bf Line Diff             & \bf 0.79                     & \bf 0.77    & \bf 0.78               \\
No Diff (Normal)     & 0.74                     & 0.72        & 0.73           \\
No Diff (Threshold - 0.01)   & 0.73                     & 0.72       & 0.73            \\
\bottomrule
\end{tabular}
\label{tab:matrix}
\end{table}

We evaluate input representations on our validation set for Defects4J 2.0. The data distribution is 72\% undetected and 
28\% detected for test matrices. The \emph{No Diff} model requires two examples per mutant, making an even more unbalanced distribution (86\% undetected, 14\% detected). Therefore, in training these models, we use a weighted loss function
that penalizes missclassfications of detected mutants more than undetected mutants. The weights are different for the \emph{Token Diff}
and \emph{Line Diff} models and the \emph{No Diff} model. 

Table \ref{tab:matrix} compares our novel input representations against the baseline Seshat model. 
\emph{Token Diff} and \emph{Line Diff} perform almost identically, with approximately a 4\%
improvement in F1 score over baseline (we use the token diff model for our other
results). Somewhat surprisingly, when the diff is not explicitly specified (in
the \emph{No Diff} models), the
model fails to reason about how code relates to tests passing or
failing This is further supported by the thresholding (in the \emph{No Diff} models)
having no effect on validation F1 score (regardless of what the threshold is from 0.01 to 0.99). 
We hypothesize that knowing the mutation applied is a key 
piece of context for accurate predictions. Both our token and line diff models have tokens that
specify the start and end of the applied operator.

\begin{table}
  \caption{\small Threshold and aggregation approaches, predicting test suites on the
    validation set. The best threshold for Seshat is 0.10; for \toolname, 0.25.
    We find that the transformer aggregation approaches have lower precision than the selected threshold approach, meaning 
    more false positives.}  
  \centering
  \begin{tabular}{l|rrrr}
  \toprule
  \bf Model             & \bf Threshold       & \bf Precision            & \bf Recall  & \bf F1         \\
  \midrule
  \multirow{5}{*}{Seshat}                &  0.10                  & 0.57        & 0.83   & 0.67                        \\
                  &  0.25                  & 0.56      &  0.85  &  0.67                  \\
                  &  0.50                  & 0.48     &  0.92    & 0.66                 \\
                  &  0.75                  & 0.52         & 0.87     & 0.65                      \\
                  &  0.90                 & 0.51         &  0.89     & 0.65                     \\
  \midrule
   \multirow{5}{*}{\toolname}             &  0.10                   & 0.76          & 0.84    & 0.80                     \\
              &  0.25                  & 0.76     &  0.84     & 0.80                \\
              &  0.50                   & 0.75     & 0.86     & 0.80                \\
              &  0.75                   & 0.74         & 0.87    & 0.80                      \\
              &  0.90                 & 0.73          & 0.88     & 0.80                     \\
  \midrule
  trans (weighted)              &  N/A                   & 0.75         &  0.85     & 0.80                     \\
  trans (unweighted)            &  N/A                    & 0.75         & 0.85    & 0.80                       \\
  \bottomrule
  \end{tabular}
\label{tab:suite}
\end{table}

We similarly evaluate aggregation strategies on the validation set, at the test
suite level (the goal of the aggregation strategies is to predict over test suites).  
Table \ref{tab:suite} shows results of all aggregation strategies we tried on the validation set. 

We find that even with the small change in F1 score between the two models for test matrix prediction, there is significant change in F1 score when it is aggregated at
the test suite level. This is due to the compounding effect of errors, as an error in any one of the tests in the test matrix can cause the whole suite to be
labeled incorrectly, making even a small difference in F1 score equate to large differences in the aggregated matrix. 

To select thresholds, we use the validation set and the F1 score followed by precision. Precision is more important than recall here, because the 
cost of a false postive is high. Specifically, a false positive means that a developer will see a mutant that is supposed to indicate test inadequacy when in reality 
their tests are adequate. We find that the best threshold for Seshat is 0.10 and the best threshold for MutationBERT is 0.25.




\subsection{RQ4: Tool Misclassifications}

\begin{table}
  \caption{\small Reasons \toolname incorrectly classifies mutants. In 71/100 cases, \toolname lacks sufficient context, while in the remaining 29/100 cases \toolname misses a 
  contextual clue.}  
  \centering
  \begin{tabular}{ll|r}
  \toprule
  \bf Category & \bf Case              & \bf Count      \\
  \midrule
  \multirow{3}{*}{Not enough context} & Helper test method & 44                \\
    & Method             & 24                           \\
    & Class             & 3                           \\
  \midrule
  \multirow{2}{*}{Missed clue} & Code                & 22             \\
                 & Method name                      & 7           \\
  \bottomrule
  \end{tabular}
\label{tab:reasons}
\vspace{-0.5cm}
\end{table}

To understand our model's limitations, we examined 100 randomly sampled examples
of \toolname misclassifications from our validation set. We categorize causes of
failures in Table~\ref{tab:reasons}. Upon inspection, we
classified each example into two high-level buckets:  \emph{Not enough context} and
\emph{Missed clue}. \emph{Not enough context} 
refers to cases where the model was missing context that even a human would need
to classify the case correctly. The large majority of our examples
(71/100) fell under this bucket. The second category consists of \emph{Missed
  clue}s, where the model missed some crucial clue to mutant behavior (29/100).

We were able to subdivide the high-level buckets into common subcategories.  For
\emph{Not enough context} these are
\emph{Helper test method}, \emph{Method} and \emph{Class}. 
\emph{Helper test method} refers to cases where the test method consists
primarily of invocations to another method. One example is as follows:

\noindent\begin{minipage}{0.9\columnwidth}
\begin{lstlisting}[language=Java]
public void testJava2DToValue() {
  checkPointsToValue(edge, plotArea);
  this.axis.setRange(0.5, 10);
  checkPointsToValue(edge, plotArea);
  ...
}
\end{lstlisting}
\end{minipage} \\
Test method \lstinline{testJava2DToValue} invokes 
helper method \lstinline{checkPointsToValue} multiple times. Without the helper method code,
\toolname lacks the context (or even knowledge of relevant test assertions) to
make an accurate prediction on any mutant.

The \emph{Method} category refers to the model lacking necessary source context.
For example:

\noindent\begin{minipage}{0.9\columnwidth}
\begin{lstlisting}[language=Java]
public <T> TypeAdapter<T> create(...)

public void testDeserializeNullField() throws IOException {
  Truck truck = truckAdapter.fromJson(...);
  ...
}
\end{lstlisting}
\end{minipage} \\
This example shows a test that invokes the \lstinline{fromJson} method, which then
invokes \lstinline{create}. Without the code for \lstinline{fromJson}, \toolname cannot reason about how
a mutant in \lstinline{create} would affect a test calling \lstinline{fromJson}.

Finally \emph{Class} refers to cases where the constructor of a class is
mutated, but the test invokes a subclass and thus is missing the subclass
constructor context. The following example shows this:

\noindent\begin{minipage}{0.9\columnwidth}
\begin{lstlisting}[language=Java]
public StrokeMap()

public void testCloning() {
  PiePlot p1 = new PiePlot(); 
  ...
}
\end{lstlisting}
\end{minipage} \\
In this example, \lstinline{testCloning} is invoking the constructor of \lstinline{PiePlot}, which is a subclass of \lstinline{StrokeMap}. Without seeing the constructor of \lstinline{PiePlot}, \toolname
 cannot understand how mutants to the \lstinline{StrokeMap} constructor affect the test.

\emph{Missed clue} is divided into  \emph{Code} and \emph{Method name}. 
\emph{Code} refers to cases where the model missed a context clue in the source
code that indicated that mutant detetion. For example:

\noindent\begin{minipage}{0.9\columnwidth}
\begin{lstlisting}[language=JavaDiff,numbers=left]
public boolean hasNext() throws IOException {
  ...
- return p != PEEKED_END_OBJECT 
-   && p != PEEKED_END_ARRAY;
+ return true && p != PEEKED_END_ARRAY;
}

public void testDoubleArrayDeserialization() {
  double[] values = gson.fromJson(...)
  assertEquals(0.0, values[0]);
  ...
}
\end{lstlisting}
\end{minipage}\\
In this example, the mutant on line 3, replaces the object check with true, but the test is only for arrays. Thus, the mutant will not be detected by the provided test, since the object check is not being tested. 
\toolname misses the correlation between the object check and the test asserts all looking at arrays.

Finally, \emph{Method name} refers to cases where the model fails to 
detect an important context clue in the method name. For example:

\noindent\begin{minipage}{0.9\columnwidth}
\begin{lstlisting}[language=JavaDiff,numbers=left]
public BufferedImage createBufferedImage(..., ChartRenderingInfo info) {
  ...
- if (info != null) {
+ if (true) {
    info.setRenderingSource(...);
  }
}

public void testDrawWithNullInfo()
\end{lstlisting}
\end{minipage} \\
This example shows a mutant that replaces a null check on \lstinline{info} with \lstinline{true}.  Since the test is a case where \lstinline{info} is null, on the mutated code, there will
be a null pointer dereference. Thus a \lstinline{NullPointerException} will be thrown and the mutant will be killed. \toolname fails to see the correlation between the test name
and the mutant applied.

\subsection{RQ5: Efficiency}

\begin{table*}
  \caption{\small Time to run Major, \toolname, and Seshat, over all mutants
    (center columns), or incorporating a confirmation check before presenting
    unkilled mutants to the user (right-hand columns).}
  \label{tab:time-savings}
\begin{tabular}{l|r|rr|rr}
\toprule
& & \multicolumn{2}{|c|}{\textbf{No Checking}} & \multicolumn{2}{|c}{\textbf{Checking}} \\\cline{3-6}
\bf Project                  & \bf Major (s) & \bf \toolname (s)           & \bf Seshat (s) & \bf \toolname (s)           & \bf Seshat (s)   \\
\midrule
commons-lang                 & 12,924           &  748 &  374 &  3324 &  5767 \\
jfreechart                   & 64,719           &  1424 &  712 &  18458 &  23838 \\
gson                         & 16,738           &  150 &  75 &  6136 &  8611 \\
commons-cli                  & 1,290            &  53 &  26 &  542 &  841 \\
jackson-core                 & 113,343          &  809 &  405 &  33035 &  52231 \\
commons-csv                  & 5,289            &  36 &  18 &  1458 &  2550 \\
\bottomrule
\end{tabular}
\end{table*}

Finally, we discuss the efficiency and performance benefits of \toolname as
compared to Major or Seshat. 
 \Cref{tab:time-savings}
shows time to run each tool, including Major, for all mutants in a project
(center column), and time to run including a confirmatory check for the
predictive techniques (right-hand columns).  

Seshat and \toolname have comparable inference time in our
experiments: 34 ms for \toolname and 17 ms for Seshat. In terms of practical
impact on a user interested in 
per-mutant prediction, the difference between 17 and 34 ms is negligible.
Meanwhile, as \Cref{tab:time-savings} shows, the time required to compute a full mutation score for a given
project is the same order of magnitude (10s of minutes), while both an order-of-magnitude
faster than Major. 


However, despite being slower than Seshat on a per-prediction basis, \toolname
still offers significant computational savings for the end-user aiming to 
improve a test suite (the original goal of mutation testing,
and consistent with its use at companies like Google and Meta). In this setting,
the user receives a list of undetected mutants to inspect and use to create new
tests. A practical application for predictive mutation testing should include a
\emph{check} of each predicted-undetected mutant before presenting the list to
the developer to filter incorrect predictions; this ensures that the tool is
presenting truly actionable information and saves the developer time and
frustration in confirming the tool's results. The right-hand-side of
\Cref{tab:time-savings} shows that because \toolname has higher precision than
Seshat (and similar recall), its predictions can be verified and thus put to use
by the developer much more quickly.


 \subsection{RQ6: Mutant Importance}

 \begin{figure}
  \begin{subfigure}{\columnwidth}
    \includegraphics[width=\columnwidth]{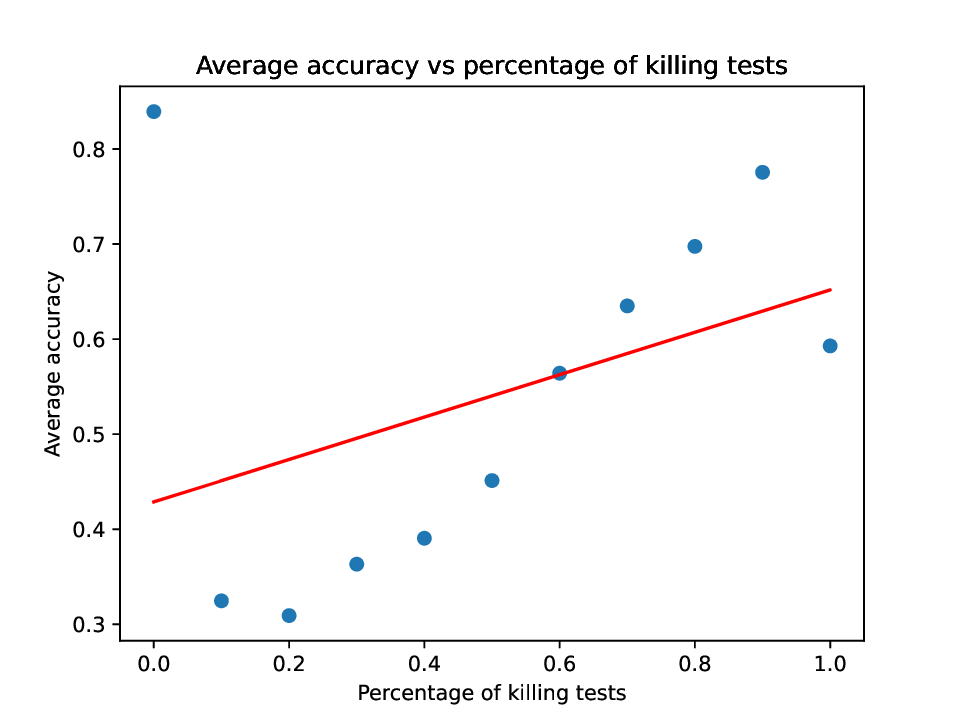}
  \caption{\small Accuracy vs. percentage of killing mutants for Seshat}
  \end{subfigure}

  \begin{subfigure}{\columnwidth}
    \includegraphics[width=\columnwidth]{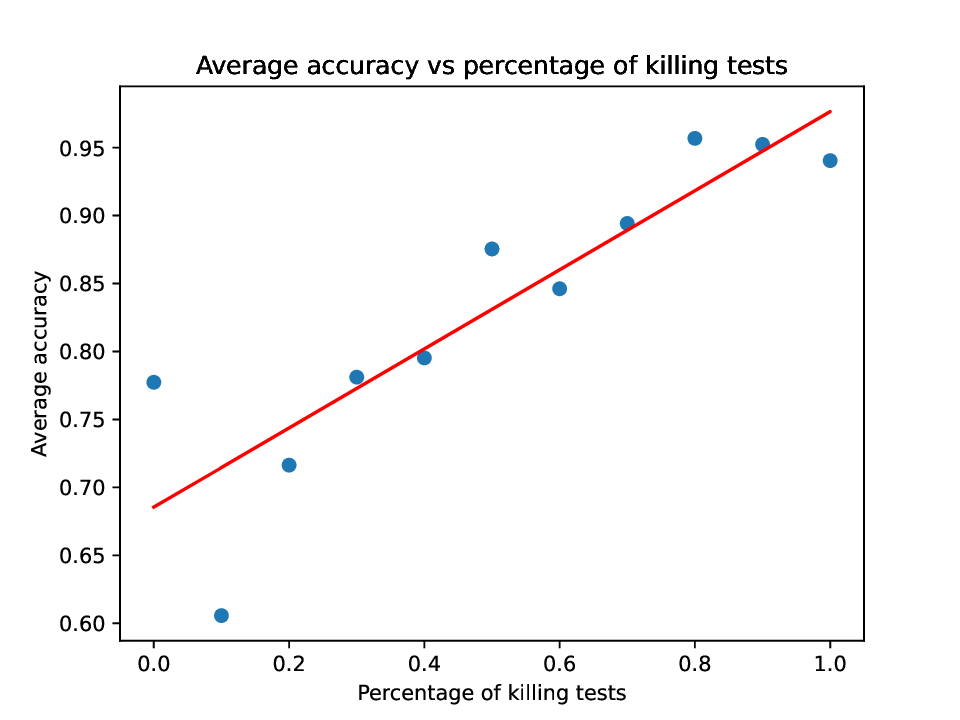}
  \caption{\small Accuracy vs. percentage of killing mutants for \toolname}
  \end{subfigure}
  \caption{Accuracy vs. percentage of killing mutants for Seshat and
    \toolname  \label{fig:acc-num-killing-mutants}}
\vspace{-3em}
  \end{figure}
 
\Cref{fig:acc-num-killing-mutants} shows model accuracy of both Seshat and \toolname with respect to percentage of detecting tests in a given mutant's test suite. 
Mutants with a high proportion of detecting tests are likely to be trivial, while mutants with few detecting tests are more likely to be interesting. We compare \toolname to Seshat in detecting trivial vs hard to detect mutants by reporting model accuracy as a function of percentage of detecting tests.
Mutants that are killed by all tests are trivial, and we hypothesize they are easier for models to detect, while mutants with fewer detecting tests are more likely to be interesting 
and more difficult for models to detect.

As expected, both approaches are less accurate at detecting mutants
that fail fewer tests. Importantly, however, \toolname outperforms Seshat considerably on harder-to-detect mutants (those failing 1\%-20\% of the test suite), by 30\%. Although Seshat is slightly more accurate at classifying mutants
that fail no tests at all (0.82 accuracy vs. 0.78), \toolname's
\emph{overall} accuracy is higher, by 17\%.  Overall, \toolname is more accurate than prior work in predicting mutant behavior, especially the hard-to-detect cases. 

\section{Discussion}

Practically, \toolname is useful for both of the core end user tasks in mutation testing: 1) as a more complete measure of testing adaquacy 
(computing mutation score)~\cite{GopinathSampleSize, offuttMutant1996} and 2) to identify undetected mutants that indicate potential inadequacies in existing testing 
efforts~\cite{BellerFacebookMutation, PetrovicMutationGoogle}.

In the classical sense, mutation testing serves to evaluate test suite quality~\cite{demillo1978hints, hamlet1977testing, jia2010analysis}. Mutation score, or the proportion
of detected mutants to total mutants, provides a powerful measure of
how well tested, including in terms of actual oracle strength, a given piece of code is. \toolname drastically reduces the amount of time needed to compute
mutation score, taking approximately 30 ms per mutant test pair,
substantially lower than the actual cost of executing a test (and
compiling mutants). The error rate of \toolname is also low, with \toolname
having below a 5\% error in predicting mutation score for both same
and cross project settings, substantially lower than Seshat.  Further note that as \Cref{tab:time-savings} shows, it is plausible
  that using MutationBERT to approximate mutation score will be faster
  (in our data, about twice as fast) as even approximating score by
  sampling as few as 10\% of mutants.  Sampling 10\% of mutants is
  likely to be no more accurate than MutationBERT~\cite{GopinathSampleSize}, and additionally
  provides \emph{no} data on mutants not sampled, while our approach provides
  a good approximation of the result for all mutants.

More recently, companies like Google~\cite{PetrovicMutationGoogle} and Facebook~\cite{BellerFacebookMutation} use mutation testing to pinpoint undetected mutants that
reveal issues with test adaquacy. \toolname substantially saves time here, as unlike Seshat, it still achieves over 60\% accuracy in predicting hard to detect mutants. When 
shown a set of undetected mutants, a developer would be able to trust \toolname's output. Even verifying the output of all mutants classified as undetected by \toolname first saves 
71\% of time when compared to regular mutation testing, significantly
more than Seshat's 57\% time savings.  We note that with very high
actual mutation scores (where examining unkilled mutants is most
useful), the time required to discover $n$ undetected mutants using
MutationBERT is likely to be \emph{much} better than with Seshat or
traditional mutation testing.

\section{Limitations and Threats}

\textbf{Limitations:}  \toolname depends on GPU availablity to efficiently make
predictions. On a CPU, \toolname takes 84 milliseconds per prediction, or 
12 mutant-test pairs per second (a far cry from the 29 mutant-test pairs per second on a GPU). Note that both these CPU and GPU times are theoretical worst cases, since these 
times were computed using a batch size of one. Many current CI pipelines are largely
CPU-based, potentially compromising practical utility.  However, cloud providers
increasingly provide GPU access; recently, GitHub actions 
announced plans to do the same for
CI.\footnote{\url{https://github.com/github/roadmap/issues/505}} Indeed, GPUs
are becoming more broadly accessible, including via idle GPU time or services
like Google Colab. Future testing approaches or any ML for SE applications are
thus increasingly realistic to deploy in practice.

\noindent\textbf{Threats to Validity:} The main internal threat to validity is our implementation of \toolname. We used widely available and popular libraries such as PyTorch and Pandas for managing 
data and building the model to help mitigate this threat.  We release our models
and implementation for inspection and extension by others.

The external threats to validity lie in our dataset of mutants and tests. We
reused the data produced by prior work on a large dataset
(Defects4J) that has been used and validated in many other studies in software
engineering. 
Since this dataset is sourced from multiple different projects, the results are more likely to generalize. 

Finally, threats to construct validity lie primarily in our evaluation metrics.
We report widely used metrics in machine learning, i.e., precision, recall and
F1 score. We also practically discuss how these metrics translate to the real world use case.

\section{Related Work}
\label{sec:related}

Several approaches have been proposed to tackle the computational cost of mutant execution, including weak-mutation, 
meta-mutation, mutation-sampling, and mutant prioritization. Offutt et. al \cite{offuttMutant1996} propose reducing the set of mutation
operators in order to prune the seach space of mutants. \citet{GopinathSampleSize} demonstrate that with a small fraction of 
mutants randomly sampled, one can easily approximate mutation score. Meta mutation \cite{untch1993mutation} combines 
multiple mutants into one larger combined mutant and executes the test
on this combined mutant.  Kaufman et. al~\cite{KaufmanFAKAJ2022} focus on computing the probability that mutants advance the
adequacy (in terms of dominating mutants) of a given test suite.

Google \cite{PetrovicMutationGoogle} and Meta \cite{BellerFacebookMutation} apply mutation testing 
\emph{only} to changed code at commit-time, and display undetected mutants as part of code review. Developers
can quickly identify potential testing gaps before code reaches
production. 
Google further uses heuristics \cite{PetrovicMutationGoogle}
to avoid mutating arid lines (lines that when mutated create
unproductive mutants, such as logging statements),
while Meta uses a learned targeted set of mutation operators
\cite{BellerFacebookMutation}.
However, even this more narrow application (just to changed code in
a commit, restricted to one mutant-per-line or a small set of operators) is expensive, requiring large amounts of idle compute
\cite{petrovicIndustrialChallenges2018}.

Approaches to reducing the cost of
mutation analysis were categorized as \textit{do smarter}, \textit{do
faster}, and \textit{do fewer} by Offutt et al.~\cite{offutt2001mutation}.
The \textit{do smarter} approaches include space-time trade-offs, weak
mutation analysis, and parallelization of mutation analysis. The \textit{do
faster} approaches include mutant schema generation  and
other methods to make mutants run faster. Finally,
\textit{do fewer} approaches include selective mutation and mutant sampling.

Recently, Predictive Mutation Testing \cite{zhang2016pmt} proposed a new means of tackling these
problems through the use of machine learning. PMT defines  a set of
features and uses these
to predict whether a given mutant is detected or not by the test
suite.  It is thus a ``do fewer'' approach in one sense, but also a
``do faster'' approach in that it reduces the runtime for a mutant
from that for test execution to the much cheaper ML inference; it is
arguably also a ``do smarter'' approach in trading off a
training run for future savings.  The original PMT approach requires costly instrumentation to collect features. Seshat \cite{kim2022predictive} achives higher accuracy with lower
overhead by exclusively using information about the source code and mutation itself (source method, test method, and mutated line).

Similar to Seshat, we also exclusively use information about the
source code and mutation itself; however we exploit CodeBERT (a model
pre-trained on source code) over the 
context of both the source and test methods along with a representation of the mutation applied. We find that this additional context is helpful in predicting the outcome 
of whether a mutant is detected or undetected, in both same-project and cross-project settings.

\section{Conclusion}

In this paper, we present \toolname, a tool for predicting both test matrices and aggregating these predictions. We perform an extensive evaluation of our model,
finding that we save 33\% of Seshat's time if a developer were to verify all mutants that either model predicted as undetected. We also outperform Seshat, the state
of the art model by 8\% \emph{F1} score in predicting test matrices and 12\% \emph{F1} score in predicting the aggregated test suite
outcome. We also achieve similar performance in the cross project setting, outperforming Seshat by 10\% \emph{F1} score in predicting test matrices and 
28\% \emph{F1} score in predicting test suites. Finally, we analyze cases where our model fails to 
classify the mutant as detected or undetected. From this analysis, we find that in the majority of cases where our model incorrectly classfies a test as detecting or failing to detect a mutant, it 
lacks sufficient context. This context often lies in test helper methods, or methods that are invoked by the test that invoke the source method.
\toolname has a relatively limited context window of 1024 tokens, so incorporating this additional information would likely
require using a large language model with larger context window sizes such as Codex. 

\section{Data Availablity}

We make all data, modeling checkpoints, and code publically available at \url{https://doi.org/10.5281/zenodo.7600371}. We include steps required to reproduce our results in the README file both from 
scratch and using our provided checkpoints.
The scripts to run our preprocessing are under \texttt{preprocessing}; scripts
to train our model are under \texttt{runtime}; and scripts to run our evaluation
on the test set are under \texttt{evaluation}. Full information on how to run each of these scripts and setup the python environment to reproduce our results is available in \texttt{README.md}.

\section{Acknowledgements}
We would like to thank the authors of Seshat for providing us with data and code for our baseline experiments. 
This work is supported in part by the US National Science Foundation, awards CCF-2129388 and CCF-1910067.

\balance

\bibliographystyle{ACM-Reference-Format}
\bibliography{bibliography}


\begin{thebibliography}{36}


\ifx \showCODEN    \undefined \def \showCODEN     #1{\unskip}     \fi
\ifx \showDOI      \undefined \def \showDOI       #1{#1}\fi
\ifx \showISBNx    \undefined \def \showISBNx     #1{\unskip}     \fi
\ifx \showISBNxiii \undefined \def \showISBNxiii  #1{\unskip}     \fi
\ifx \showISSN     \undefined \def \showISSN      #1{\unskip}     \fi
\ifx \showLCCN     \undefined \def \showLCCN      #1{\unskip}     \fi
\ifx \shownote     \undefined \def \shownote      #1{#1}          \fi
\ifx \showarticletitle \undefined \def \showarticletitle #1{#1}   \fi
\ifx \showURL      \undefined \def \showURL       {\relax}        \fi
\providecommand\bibfield[2]{#2}
\providecommand\bibinfo[2]{#2}
\providecommand\natexlab[1]{#1}
\providecommand\showeprint[2][]{arXiv:#2}

\bibitem[Aghamohammadi and Mirian-Hosseinabadi(2020)]%
        {AghamohammadiThreatsPMT}
\bibfield{author}{\bibinfo{person}{Alireza Aghamohammadi} {and}
  \bibinfo{person}{Seyed-Hassan Mirian-Hosseinabadi}.}
  \bibinfo{year}{2020}\natexlab{}.
\newblock \showarticletitle{The Threat to the Validity of Predictive Mutation
  Testing: The Impact of Uncovered Mutants}.
\newblock \bibinfo{journal}{\emph{CoRR}}  \bibinfo{volume}{abs/2005.11532}
  (\bibinfo{year}{2020}).
\newblock


\bibitem[Ahmad et~al\mbox{.}(2021)]%
        {ahmad-etal-2021-unified}
\bibfield{author}{\bibinfo{person}{Wasi Ahmad}, \bibinfo{person}{Saikat
  Chakraborty}, \bibinfo{person}{Baishakhi Ray}, {and} \bibinfo{person}{Kai-Wei
  Chang}.} \bibinfo{year}{2021}\natexlab{}.
\newblock \showarticletitle{{Unified Pre-training for Program Understanding and
  Generation}}. In \bibinfo{booktitle}{\emph{North American Chapter of the
  Association for Computational Linguistics: Human Language Technologies}}
  \emph{(\bibinfo{series}{NAACL-HLT '21})}. \bibinfo{pages}{2655--2668}.
\newblock


\bibitem[Ahmed and Devanbu(2023)]%
        {Ahmed2022FewshotTL}
\bibfield{author}{\bibinfo{person}{Toufique Ahmed} {and}
  \bibinfo{person}{Premkumar Devanbu}.} \bibinfo{year}{2023}\natexlab{}.
\newblock \showarticletitle{Few-Shot Training LLMs for Project-Specific
  Code-Summarization}. In \bibinfo{booktitle}{\emph{Automated Software
  Engineering}} \emph{(\bibinfo{series}{ASE '23})}. Article
  \bibinfo{articleno}{177}, \bibinfo{numpages}{5}~pages.
\newblock


\bibitem[Beller et~al\mbox{.}(2021)]%
        {BellerFacebookMutation}
\bibfield{author}{\bibinfo{person}{Moritz Beller}, \bibinfo{person}{Chu{-}Pan
  Wong}, \bibinfo{person}{Johannes Bader}, \bibinfo{person}{Andrew Scott},
  \bibinfo{person}{Mateusz Machalica}, \bibinfo{person}{Satish Chandra}, {and}
  \bibinfo{person}{Erik Meijer}.} \bibinfo{year}{2021}\natexlab{}.
\newblock \showarticletitle{What It Would Take to Use Mutation Testing in
  Industry - {A} Study at Facebook}. In \bibinfo{booktitle}{\emph{International
  Conference on Software Engineering: Software Engineering in Practice}}
  \emph{(\bibinfo{series}{ICSE '18})}. \bibinfo{publisher}{{IEEE}},
  \bibinfo{pages}{268--277}.
\newblock


\bibitem[Bokaei and Keyvanpour(2019)]%
        {BokaeiChallenges2019}
\bibfield{author}{\bibinfo{person}{N.~N. Bokaei} {and} \bibinfo{person}{M.~R.
  Keyvanpour}.} \bibinfo{year}{2019}\natexlab{}.
\newblock \showarticletitle{A Comparative Study of {Whole Issues} and
  {Challenges} in {Mutation Testing}}. In \bibinfo{booktitle}{\emph{Conference
  on Knowledge Based Engineering and Innovation}} \emph{(\bibinfo{series}{KBEI
  '19})}. \bibinfo{pages}{745--754}.
\newblock


\bibitem[Coles et~al\mbox{.}(2016)]%
        {colesPIT2016}
\bibfield{author}{\bibinfo{person}{Henry Coles}, \bibinfo{person}{Thomas
  Laurent}, \bibinfo{person}{Christopher Henard}, \bibinfo{person}{Mike
  Papadakis}, {and} \bibinfo{person}{Anthony Ventresque}.}
  \bibinfo{year}{2016}\natexlab{}.
\newblock \showarticletitle{PIT: A Practical Mutation Testing Tool for {Java}
  (Demo)}. In \bibinfo{booktitle}{\emph{International Symposium on Software
  Testing and Analysis}} \emph{(\bibinfo{series}{ISSTA '16})}.
  \bibinfo{pages}{449--452}.
\newblock


\bibitem[DeMillo et~al\mbox{.}(1978)]%
        {demillo1978hints}
\bibfield{author}{\bibinfo{person}{R.~A. DeMillo}, \bibinfo{person}{R.~J.
  Lipton}, {and} \bibinfo{person}{F.~G. Sayward}.}
  \bibinfo{year}{1978}\natexlab{}.
\newblock \showarticletitle{Hints on {Test} Data Selection: Help for the
  Practicing Programmer}.
\newblock \bibinfo{journal}{\emph{IEEE Computer}} \bibinfo{volume}{11},
  \bibinfo{number}{4} (\bibinfo{date}{Apr} \bibinfo{year}{1978}),
  \bibinfo{pages}{34--41}.
\newblock


\bibitem[Feng et~al\mbox{.}(2020)]%
        {feng-etal-2020-codebert}
\bibfield{author}{\bibinfo{person}{Zhangyin Feng}, \bibinfo{person}{Daya Guo},
  \bibinfo{person}{Duyu Tang}, \bibinfo{person}{Nan Duan},
  \bibinfo{person}{Xiaocheng Feng}, \bibinfo{person}{Ming Gong},
  \bibinfo{person}{Linjun Shou}, \bibinfo{person}{Bing Qin},
  \bibinfo{person}{Ting Liu}, \bibinfo{person}{Daxin Jiang}, {and}
  \bibinfo{person}{Ming Zhou}.} \bibinfo{year}{2020}\natexlab{}.
\newblock \showarticletitle{{CodeBERT}: A Pre-Trained Model for Programming and
  Natural Languages}. In \bibinfo{booktitle}{\emph{Findings of the Association
  for Computational Linguistics: EMNLP}} \emph{(\bibinfo{series}{EMNLP '20})}.
  \bibinfo{pages}{1536--1547}.
\newblock


\bibitem[Gopinath et~al\mbox{.}(2015)]%
        {GopinathSampleSize}
\bibfield{author}{\bibinfo{person}{Rahul Gopinath}, \bibinfo{person}{Amin
  Alipour}, \bibinfo{person}{Iftekhar Ahmed}, \bibinfo{person}{Carlos Jensen},
  {and} \bibinfo{person}{Alex Groce}.} \bibinfo{year}{2015}\natexlab{}.
\newblock \showarticletitle{How Hard Does Mutation Analysis Have to Be,
  Anyway?}. In \bibinfo{booktitle}{\emph{Software Reliability Engineering}}.
  \bibinfo{pages}{216--227}.
\newblock


\bibitem[Groce et~al\mbox{.}(2018)]%
        {universalMutator}
\bibfield{author}{\bibinfo{person}{Alex Groce}, \bibinfo{person}{Josie Holmes},
  \bibinfo{person}{Darko Marinov}, \bibinfo{person}{August Shi}, {and}
  \bibinfo{person}{Lingming Zhang}.} \bibinfo{year}{2018}\natexlab{}.
\newblock \showarticletitle{An Extensible, {Regular-Expression}-Based Tool for
  Multi-Language Mutant Generation}. In \bibinfo{booktitle}{\emph{International
  Conference on Software Engineering}} \emph{(\bibinfo{series}{ICSE '18})}.
  \bibinfo{pages}{25--28}.
\newblock


\bibitem[Groce et~al\mbox{.}(2022)]%
        {icseseip22}
\bibfield{author}{\bibinfo{person}{Alex Groce}, \bibinfo{person}{Kush Jain},
  \bibinfo{person}{Rijnard van Tonder}, \bibinfo{person}{Goutamkumar~Tulajappa
  Kalburgi}, {and} \bibinfo{person}{Claire Le~Goues}.}
  \bibinfo{year}{2022}\natexlab{}.
\newblock \showarticletitle{Looking for {Lacunae} in {Bitcoin Core's} {Fuzzing
  Efforts}}. In \bibinfo{booktitle}{\emph{International Conference on Software
  Engineering: Software Engineering in Practice}} \emph{(\bibinfo{series}{ICSE
  '22})}.
\newblock


\bibitem[Hamlet(1977)]%
        {hamlet1977testing}
\bibfield{author}{\bibinfo{person}{R.G. Hamlet}.}
  \bibinfo{year}{1977}\natexlab{}.
\newblock \showarticletitle{Testing Programs with the Aid of a Compiler}.
\newblock \bibinfo{journal}{\emph{IEEE Transactions on Software Engineering}}
  \bibinfo{volume}{SE-3}, \bibinfo{number}{4} (\bibinfo{year}{1977}),
  \bibinfo{pages}{279--290}.
\newblock


\bibitem[Hellendoorn and Devanbu(2017)]%
        {hellendoorn2017deep}
\bibfield{author}{\bibinfo{person}{Vincent~J Hellendoorn} {and}
  \bibinfo{person}{Premkumar Devanbu}.} \bibinfo{year}{2017}\natexlab{}.
\newblock \showarticletitle{Are {Deep Neural Networks} the Best Choice for
  Modeling Source Code?}. In \bibinfo{booktitle}{\emph{Joint Meeting of the
  European Software Engineering Conference and the Symposium on the Foundations
  of Software Engineering}} \emph{(\bibinfo{series}{ESEC/FSE '17})}.
  \bibinfo{pages}{763--773}.
\newblock


\bibitem[Howden(1982)]%
        {HowdenWeak1982}
\bibfield{author}{\bibinfo{person}{W.E. Howden}.}
  \bibinfo{year}{1982}\natexlab{}.
\newblock \showarticletitle{Weak Mutation Testing and Completeness of Test
  Sets}.
\newblock \bibinfo{journal}{\emph{IEEE Transactions on Software Engineering}}
  \bibinfo{volume}{SE-8}, \bibinfo{number}{4} (\bibinfo{year}{1982}),
  \bibinfo{pages}{371--379}.
\newblock


\bibitem[Jia and Harman(2010)]%
        {jia2010analysis}
\bibfield{author}{\bibinfo{person}{Yue Jia} {and} \bibinfo{person}{Mark
  Harman}.} \bibinfo{year}{2010}\natexlab{}.
\newblock \showarticletitle{An analysis and survey of the development of
  mutation testing}.
\newblock \bibinfo{journal}{\emph{IEEE Transactions on Software Engineering}}
  \bibinfo{volume}{37}, \bibinfo{number}{5} (\bibinfo{year}{2010}),
  \bibinfo{pages}{649--678}.
\newblock


\bibitem[Just(2014)]%
        {just2014Major}
\bibfield{author}{\bibinfo{person}{Ren{\'{e}} Just}.}
  \bibinfo{year}{2014}\natexlab{}.
\newblock \showarticletitle{The Major Mutation Framework: Efficient and
  Scalable Mutation Analysis for Java}. In
  \bibinfo{booktitle}{\emph{International Symposium on Software Testing and
  Analysis}} \emph{(\bibinfo{series}{ISSTA '14})}.
  \bibinfo{publisher}{Association for Computing Machinery},
  \bibinfo{pages}{433--436}.
\newblock


\bibitem[Just et~al\mbox{.}(2014)]%
        {just2014mutants}
\bibfield{author}{\bibinfo{person}{Ren{\'e} Just}, \bibinfo{person}{Darioush
  Jalali}, \bibinfo{person}{Laura Inozemtseva}, \bibinfo{person}{Michael~D
  Ernst}, \bibinfo{person}{Reid Holmes}, {and} \bibinfo{person}{Gordon
  Fraser}.} \bibinfo{year}{2014}\natexlab{}.
\newblock \showarticletitle{Are Mutants a Valid Substitute for Real Faults in
  Software Testing?}. In \bibinfo{booktitle}{\emph{Symposium on Foundations of
  Software Engineering}} \emph{(\bibinfo{series}{FSE '14})}.
  \bibinfo{pages}{654--665}.
\newblock


\bibitem[Just et~al\mbox{.}(2017)]%
        {MutantUtilityContext}
\bibfield{author}{\bibinfo{person}{Ren\'{e} Just}, \bibinfo{person}{Bob Kurtz},
  {and} \bibinfo{person}{Paul Ammann}.} \bibinfo{year}{2017}\natexlab{}.
\newblock \showarticletitle{Inferring {Mutant} {Utility} from {Program}
  {Context}}. In \bibinfo{booktitle}{\emph{ACM SIGSOFT International Symposium
  on Software Testing and Analysis}} \emph{(\bibinfo{series}{ISSTA '17})}.
  \bibinfo{pages}{284--294}.
\newblock


\bibitem[Kaufman et~al\mbox{.}(2022)]%
        {KaufmanFAKAJ2022}
\bibfield{author}{\bibinfo{person}{Samuel~J. Kaufman}, \bibinfo{person}{Ryan
  Featherman}, \bibinfo{person}{Justin Alvin}, \bibinfo{person}{Bob Kurtz},
  \bibinfo{person}{Paul Ammann}, {and} \bibinfo{person}{Ren{\'e} Just}.}
  \bibinfo{year}{2022}\natexlab{}.
\newblock \showarticletitle{Prioritizing Mutants to Guide Mutation Testing}. In
  \bibinfo{booktitle}{\emph{International Conference on Software Engineering}}
  \emph{(\bibinfo{series}{ICSE '22})}.
\newblock


\bibitem[Kim et~al\mbox{.}(2022)]%
        {kim2022predictive}
\bibfield{author}{\bibinfo{person}{Jinhan Kim}, \bibinfo{person}{Juyoung Jeon},
  \bibinfo{person}{Shin Hong}, {and} \bibinfo{person}{Shin Yoo}.}
  \bibinfo{year}{2022}\natexlab{}.
\newblock \showarticletitle{Predictive Mutation Analysis via the {Natural
  Language} Channel in Source Code}.
\newblock \bibinfo{journal}{\emph{ACM Transactions on Software Engineering
  Methodology}} \bibinfo{volume}{31}, \bibinfo{number}{4}, Article
  \bibinfo{articleno}{73} (\bibinfo{year}{2022}).
\newblock


\bibitem[Ma et~al\mbox{.}(2022)]%
        {maGraphCode2Vec2022}
\bibfield{author}{\bibinfo{person}{Wei Ma}, \bibinfo{person}{Mengjie Zhao},
  \bibinfo{person}{Ezekiel Soremekun}, \bibinfo{person}{Qiang Hu},
  \bibinfo{person}{Jie~M. Zhang}, \bibinfo{person}{Mike Papadakis},
  \bibinfo{person}{Maxime Cordy}, \bibinfo{person}{Xiaofei Xie}, {and}
  \bibinfo{person}{Yves~Le Traon}.} \bibinfo{year}{2022}\natexlab{}.
\newblock \showarticletitle{{GraphCode2Vec}: Generic Code Embedding via Lexical
  and Program Dependence Analyses}. In \bibinfo{booktitle}{\emph{Mining
  Software Repositories}} \emph{(\bibinfo{series}{MSR '22})}.
  \bibinfo{pages}{524--536}.
\newblock


\bibitem[Mao et~al\mbox{.}(2019)]%
        {mao2019crossproject}
\bibfield{author}{\bibinfo{person}{Dongyu Mao}, \bibinfo{person}{Lingchao
  Chen}, {and} \bibinfo{person}{Lingming Zhang}.}
  \bibinfo{year}{2019}\natexlab{}.
\newblock \showarticletitle{An Extensive Study on {Cross-Project} Predictive
  Mutation Testing}. In \bibinfo{booktitle}{\emph{Software Testing, Validation
  and Verification}} \emph{(\bibinfo{series}{ICST '19})}.
  \bibinfo{pages}{160--171}.
\newblock


\bibitem[Offutt et~al\mbox{.}(1996)]%
        {offuttMutant1996}
\bibfield{author}{\bibinfo{person}{A.~Jefferson Offutt}, \bibinfo{person}{Ammei
  Lee}, \bibinfo{person}{Gregg Rothermel}, \bibinfo{person}{Roland~H. Untch},
  {and} \bibinfo{person}{Christian Zapf}.} \bibinfo{year}{1996}\natexlab{}.
\newblock \showarticletitle{An Experimental Determination of Sufficient Mutant
  Operators}.
\newblock \bibinfo{journal}{\emph{ACM Transactions on Software Engineering
  Methodology}} \bibinfo{volume}{5}, \bibinfo{number}{2}
  (\bibinfo{year}{1996}), \bibinfo{pages}{99--118}.
\newblock


\bibitem[Offutt and Untch(2001)]%
        {offutt2001mutation}
\bibfield{author}{\bibinfo{person}{A.~Jefferson Offutt} {and}
  \bibinfo{person}{Roland~H. Untch}.} \bibinfo{year}{2001}\natexlab{}.
\newblock \showarticletitle{Mutation 2000: Uniting the Orthogonal}.
\newblock In \bibinfo{booktitle}{\emph{Mutation Testing for the New Century}}.
  \bibinfo{publisher}{Springer}, \bibinfo{pages}{34--44}.
\newblock


\bibitem[Papadakis et~al\mbox{.}(2018)]%
        {papadakis2018mutation}
\bibfield{author}{\bibinfo{person}{Mike Papadakis}, \bibinfo{person}{Donghwan
  Shin}, \bibinfo{person}{Shin Yoo}, {and} \bibinfo{person}{Doo-Hwan Bae}.}
  \bibinfo{year}{2018}\natexlab{}.
\newblock \showarticletitle{Are Mutation Scores Correlated with Real Fault
  Detection? A Large Scale Empirical Study on the Relationship between Mutants
  and Real Faults}. In \bibinfo{booktitle}{\emph{International Conference on
  Software Engineering}} \emph{(\bibinfo{series}{ICSE '18})}.
  \bibinfo{pages}{537--548}.
\newblock


\bibitem[Petrovic and Ivankovic(2018)]%
        {PetrovicMutationGoogle}
\bibfield{author}{\bibinfo{person}{Goran Petrovic} {and} \bibinfo{person}{Marko
  Ivankovic}.} \bibinfo{year}{2018}\natexlab{}.
\newblock \showarticletitle{State of Mutation Testing at {Google}}. In
  \bibinfo{booktitle}{\emph{International Conference on Software Engineering:
  Software Engineering in Practice}} \emph{(\bibinfo{series}{ICSE '18})}.
  \bibinfo{pages}{163--171}.
\newblock


\bibitem[Petrovic et~al\mbox{.}(2018)]%
        {petrovicIndustrialChallenges2018}
\bibfield{author}{\bibinfo{person}{Goran Petrovic}, \bibinfo{person}{Marko
  Ivankovic}, \bibinfo{person}{Bob Kurtz}, \bibinfo{person}{Paul Ammann}, {and}
  \bibinfo{person}{René Just}.} \bibinfo{year}{2018}\natexlab{}.
\newblock \showarticletitle{An Industrial Application of Mutation Testing:
  Lessons, Challenges, and Research Directions}. In
  \bibinfo{booktitle}{\emph{Software Testing, Verification and Validation
  Workshops}} \emph{(\bibinfo{series}{ICSTW '18})}. \bibinfo{pages}{47--53}.
\newblock


\bibitem[Popel and Bojar(2018)]%
        {PopelTrainingTips}
\bibfield{author}{\bibinfo{person}{Martin Popel} {and} \bibinfo{person}{Ondrej
  Bojar}.} \bibinfo{year}{2018}\natexlab{}.
\newblock \showarticletitle{{Training Tips for the {Transformer Model}}}.
\newblock \bibinfo{journal}{\emph{CoRR}}  \bibinfo{volume}{abs/1804.00247}
  (\bibinfo{year}{2018}).
\newblock


\bibitem[Sennrich et~al\mbox{.}(2016)]%
        {sennrich-etal-2016-neural}
\bibfield{author}{\bibinfo{person}{Rico Sennrich}, \bibinfo{person}{Barry
  Haddow}, {and} \bibinfo{person}{Alexandra Birch}.}
  \bibinfo{year}{2016}\natexlab{}.
\newblock \showarticletitle{Neural Machine Translation of Rare Words with
  Subword Units}. In \bibinfo{booktitle}{\emph{Association for Computational
  Linguistics}} \emph{(\bibinfo{series}{ACL '16})}.
  \bibinfo{pages}{1715--1725}.
\newblock


\bibitem[Svyatkovskiy et~al\mbox{.}(2022)]%
        {svyatkovskiy2021mergebert}
\bibfield{author}{\bibinfo{person}{Alexey Svyatkovskiy}, \bibinfo{person}{Sarah
  Fakhoury}, \bibinfo{person}{Negar Ghorbani}, \bibinfo{person}{Todd
  Mytkowicz}, \bibinfo{person}{Elizabeth Dinella}, \bibinfo{person}{Christian
  Bird}, \bibinfo{person}{Jinu Jang}, \bibinfo{person}{Neel Sundaresan}, {and}
  \bibinfo{person}{Shuvendu~K. Lahiri}.} \bibinfo{year}{2022}\natexlab{}.
\newblock \showarticletitle{Program {Merge} {Conflict} {Resolution} via
  {Neural} {Transformers}}. In \bibinfo{booktitle}{\emph{Symposium on the
  Foundations of Software Engineering}} \emph{(\bibinfo{series}{FSE '22})}.
  \bibinfo{pages}{822--833}.
\newblock


\bibitem[Untch et~al\mbox{.}(1993)]%
        {untch1993mutation}
\bibfield{author}{\bibinfo{person}{Roland~H. Untch},
  \bibinfo{person}{A.~Jefferson Offutt}, {and} \bibinfo{person}{Mary~Jean
  Harrold}.} \bibinfo{year}{1993}\natexlab{}.
\newblock \showarticletitle{Mutation Analysis Using {Mutant} Schemata}.
\newblock \bibinfo{journal}{\emph{ACM SIGSOFT Software Engineering Notes}}
  \bibinfo{volume}{18}, \bibinfo{number}{3} (\bibinfo{year}{1993}),
  \bibinfo{pages}{139--148}.
\newblock


\bibitem[Vaswani et~al\mbox{.}(2017)]%
        {vaswani2017attention}
\bibfield{author}{\bibinfo{person}{Ashish Vaswani}, \bibinfo{person}{Noam
  Shazeer}, \bibinfo{person}{Niki Parmar}, \bibinfo{person}{Jakob Uszkoreit},
  \bibinfo{person}{Llion Jones}, \bibinfo{person}{Aidan~N. Gomez},
  \bibinfo{person}{{\L}ukasz Kaiser}, {and} \bibinfo{person}{Illia
  Polosukhin}.} \bibinfo{year}{2017}\natexlab{}.
\newblock \showarticletitle{Attention is all you need}.
\newblock \bibinfo{journal}{\emph{Advances in Neural Information Processing
  Systems}}  \bibinfo{volume}{30} (\bibinfo{year}{2017}).
\newblock


\bibitem[Wang et~al\mbox{.}(2020)]%
        {WenhanCodeClone2020}
\bibfield{author}{\bibinfo{person}{Wenhan Wang}, \bibinfo{person}{Ge Li},
  \bibinfo{person}{Bo Ma}, \bibinfo{person}{Xin Xia}, {and}
  \bibinfo{person}{Zhi Jin}.} \bibinfo{year}{2020}\natexlab{}.
\newblock \showarticletitle{Detecting Code Clones with Graph Neural Network and
  Flow-Augmented Abstract Syntax Tree}.
\newblock \bibinfo{journal}{\emph{CoRR}}  \bibinfo{volume}{abs/2002.08653}
  (\bibinfo{year}{2020}).
\newblock


\bibitem[Wang et~al\mbox{.}(2021)]%
        {wang-etal-2021-codet5}
\bibfield{author}{\bibinfo{person}{Yue Wang}, \bibinfo{person}{Weishi Wang},
  \bibinfo{person}{Shafiq Joty}, {and} \bibinfo{person}{Steven~C.H. Hoi}.}
  \bibinfo{year}{2021}\natexlab{}.
\newblock \showarticletitle{{CodeT5}: Identifier-aware Unified Pre-trained
  Encoder-Decoder Models for Code Understanding and Generation}. In
  \bibinfo{booktitle}{\emph{Empirical Methods in Natural Language Processing}}
  \emph{(\bibinfo{series}{EMNLP '21})}. \bibinfo{publisher}{Association for
  Computational Linguistics}, \bibinfo{pages}{8696--8708}.
\newblock


\bibitem[Yasunaga and Liang(2020)]%
        {YasunagaGraphBased2020}
\bibfield{author}{\bibinfo{person}{Michihiro Yasunaga} {and}
  \bibinfo{person}{Percy Liang}.} \bibinfo{year}{2020}\natexlab{}.
\newblock \showarticletitle{Graph-Based, Self-Supervised Program Repair from
  Diagnostic Feedback}. In \bibinfo{booktitle}{\emph{International Conference
  on Machine Learning}} \emph{(\bibinfo{series}{ICML'20})}. Article
  \bibinfo{articleno}{1001}.
\newblock


\bibitem[Zhang et~al\mbox{.}(2016)]%
        {zhang2016pmt}
\bibfield{author}{\bibinfo{person}{Jie Zhang}, \bibinfo{person}{Ziyi Wang},
  \bibinfo{person}{Lingming Zhang}, \bibinfo{person}{Dan Hao},
  \bibinfo{person}{Lei Zang}, \bibinfo{person}{Shiyang Cheng}, {and}
  \bibinfo{person}{Lu Zhang}.} \bibinfo{year}{2016}\natexlab{}.
\newblock \showarticletitle{Predictive Mutation Testing}. In
  \bibinfo{booktitle}{\emph{International Symposium on Software Testing and
  Analysis}} \emph{(\bibinfo{series}{ISSTA '16})}. \bibinfo{pages}{342--353}.
\newblock


\end{thebibliography}

\balance

\end{document}